\documentclass[aps,prl,superscriptaddress,twocolumn,showpacs,preprintnumbers,nofootinbib]{revtex4-1}
\usepackage{amsfonts}
\usepackage{amssymb,amsmath}
\usepackage{mathrsfs}
\usepackage{amsthm}
\usepackage{newlfont}
\usepackage{graphicx}

\newcommand{\usb}{\affiliation{Departamento de F\'{\i}sica, Secci\'{o}n de Fen\'{o}menos \'{O}pticos, Universidad Sim\'{o}n Bol\'{\i}var,Apartado Postal 89000, Caracas 1080-A, Venezuela.}}
\newcommand{\ivic}{\affiliation{Centro de F\'{\i}sica, Instituto Venezolano de Investigaciones Cient\'{\i}ficas, Apartado 20632 Caracas 1020-A, Venezuela.}}

\begin{document}
\begin{flushright} ${}$\\[-40pt] $\scriptstyle \mathrm SB/F/432-14$ \\[0pt]
\end{flushright}
\title{The force density and the kinetic energy-momentum tensor of
electromagnetic fields in matter} 
\author{Rodrigo Medina}\email[]{rmedina@ivic.gob.ve}\ivic
\author{J Stephany}\email[]{stephany@usb.ve}\usb

\begin{abstract}
We determine the invariant expression of the force density that the
electromagnetic field exerts on dipolar matter and construct the non-symmetric energy-momentum tensor of the electromagnetic field in matter which is consistent with that force and with Maxwell equations.  We recover Minkowski's expression for the momentum density.
We use our results to discuss momentum exchange of an electromagnetic
wave-packet which falls into a dielectric block.  In particular we show
that the wave-packet pulls the block when it enters and drags it when it
leaves. The usual form of the center of mass motion theorem  does not hold for this system but a modified version of the theorem which includes a spin contribution is shown to be satisfied.
\end{abstract}
\maketitle 
\thispagestyle{empty} 

\newpage
\newpage
The Abraham-Minkowski controversy on the momentum of the electromagnetic field in matter has a long story.  In 1908 Minkowski \cite{Minkowski1908} proposed a
non-symmetric energy-momentum tensor. For photons with energy $E$ it implies a momentum  $nE/c$ with $n$ the refraction index.  A year later Abraham \cite{Abraham1910}, arguing that angular momentum conservation
requires the tensor to be symmetric,made a proposal for
which photon's momentum is $E/nc$ . Since then many theoretical and
experimental arguments have been exposed which favor one
or the other tensor. Reviews of the  controversy
can be found in \cite{Brevik1979,Pfeifer2007,MilonniBoyd2010,BarnettLoudon2010}.
Abraham's premise of symmetry was long ago overruled by the discovery of spin,
but arguments apparently independent  appeared to back his proposal, notably
one based in the the so called
center of mass motion theorem (CMMT), which states that the center of mass
of an isolated system moves with constant velocity \cite{Balazs1953}.  The argument says that since Minkowski's momentum in matter is greater than in vacuum, photons crossing a
dielectric block will  pull the block instead of pushing it and the CMMT will be
violated. As we discuss below, the CMMT  only holds  \cite{MedandSa,MedandSf} for systems for which  the energy-momentum tensor $T_{\mu\nu}$ is symmetric, that is in the absence of spin. This an other misunderstandings related to the CMMT had  populated the literature on the subject with constructions which depart from standard Lorentz-Maxwell elecrodynamics. Among them the hidden momentum hypothesis \cite{SJ1967,CvV1968} and the use of force densities which are not obtained from the microscopic Lorentz force \cite{Brevik1979,Pfeifer2007,GriD2012}. In this letter we show that none of this is necessary and that Balzacs argument is wrong. This is done by computing the correct energy momentum tensor of the electromagnetig field in matter and then showing by an explicit computation that for an electromagnetic wave which falls on a dielectric block CMMT does not hold but and improved version of the theorem which includes spin is satisfied.

To discuss the CMMT  consider an isolated, localized system  with a non-symmetric conserved energy-momentum tensor $\partial_\nu T^{\mu\nu}=0$ and a non vanishing local spin density $S^{\mu\nu\alpha}$. The total energy $U=\int T^{00}dV$ and the total momentum
$p^i = c^{-1}\int T^{i0}dV$ are conserved. The current density of the orbital angular momentum, $L^{\mu\nu\alpha} = x^\mu T^{\nu\alpha}-x^\nu T^{\mu\alpha} \ $ is not conserved
\begin{equation}
\label{orbital-eq}
\partial_\alpha L^{\mu\nu\alpha} = T^{\nu\mu}-T^{\mu\nu} \ \ .
\end{equation}
Imposing instead the conservation of the total angular momentum current density $J^{\mu\nu\alpha}= L^{\mu\nu\alpha}+S^{\mu\nu\alpha}$, one has  \cite{Papapetrou1949,MedandSd},
\begin{equation}
\label{spin-eq}
\partial_\alpha S^{\mu\nu\alpha} = T^{\mu\nu}-T^{\nu\mu} \ \ .
\end{equation}
Define the  center of mass by
\begin{equation}
\label{CenterMass}
X^i_T = \frac{1}{U}\int x^i T^{00}\,dV\ \  .
\end{equation}
In the case when there is no spin, $T^{\mu\nu}$ is symmetric and the orbital angular momentum $L^{\mu\nu}=c^{-1}\int L^{\mu\nu 0}\,dV$ is conserved. Then, it is easy to see that the center of mass  moves with velocity $c^2 p^i/U$. For the non-symmetric $T^{\mu\nu}$ we are considering it is also easy to see that
\begin{equation}
\label{CenterMass-velocity}
\dot{X}^i_T = \frac{c}{U}\int T^{0i}\,dV\ \ .
\end{equation}
In (\ref{CenterMass-velocity}) appears  the energy current
density and not the momentum density. The CMMT is not  obtained. This is a consequence of the non-vanishing spin of the system. To see why define the spin matrix $S^{\mu\nu}= c^{-1}\int S^{\mu\nu 0}\,dV$ and consider the quantity  
\begin{equation}
\label{spincenter}
X^i_S=-\frac{c}{U}S^{0i}\ .
\end{equation} 
From the conservation of the total angular momentum it follows directly that,
\begin{eqnarray}
\label{CMSMT}
\frac{d}{dt}X^i_S&=&\frac{c}{U} \frac{d}{dt} L^{0i}= \frac{1}{U}\frac{d}{dt}\int \big[x^0 T^{i0}-x^iT^{00}\big]dv\nonumber\\
&=&\frac{c^2 p^i}{U}-\frac{d}{dt}X^i_T\ .
\end{eqnarray}
The center of mass and spin defined by 
\begin{equation}
 X^i_\Theta=X^i_T+X^i_S
\end{equation}
moves with constant velocity $\dot{X}^i_\Theta = c^2 p^i/U $. It is worth noting \cite{MedandSe}, that $X^i_\Theta$ corresponds to the center of mass computed from the symmetric Belifante-Rosenfeld tensor \cite{BelF1939,Rosenfeld1940} which is a combination of the energy-momentum tensor and the spin density. In the literature  the Belinfante-Rosenfeld tensor is frequently considered as an improved symmetrized energy-momentum tensor but our discussion shows  that this interpretation, at least from the mechanical point of view is wrong. Spin and energy-momentun should be distinguished.   To illustrate this  consider a magnet with total magnetic moment different of zero. The spatial part of the spin density is proportional to magnetization. In the Einstein-de Haas experiment which is  used routinely to measure the gyromagnetic radio \cite{BarKen1952}, spin is converted in orbital angular momentum. This proccess is described by equation (\ref{spin-eq}) and provides an example where the total energy-momentum 
tensor clearly cannot be symmetric. For another interesting example see Ref.\cite{MedandSa}.

We now turn to the computation of the energy-momentum tensor of the electromagnetic field in matter. Contrary to the common belief this can be done unequivocally. The force density on matter is in principle an observable quantity and on theoretical grounds it expression should be deduced from the microscopic Lorentz force. In absence of other interactions the divergence of the energy-momentum of matter is given by this force density. Conservation of momentum then requieres that Newton third law holds implying that the divergence of the electromagnetic energy-momentum tensor should be minus the force density.   Consequently the key points to solve our problem are  to identify the correct density of force which is deduced from the microscopic Lorentz force and to use the action-reaction principle between matter and field. As we show below is also important  to take full advantage of the relativistic character of the polarization tensor. So, let us consider a matter system with free charge and current densities 
$\rho$ and $\mathbf{j} $ , polarization  $\mathbf{P}$ and  magnetization $\mathbf{M}$.
The bound charge density is $\rho_\mathrm{b}=-\mathbf{\nabla} \cdot\mathbf{P}$,
the bound current density  is $\frac{\partial \mathbf{P}}{\partial t}$ and the
magnetization current density is  $\mathbf{j}_{\mathrm{M}}=c\mathbf{\nabla \times M}$. In the
surface of a piece of material there are a surface  density of bound charge
$\mathbf{P}\cdot\hat{\mathbf{n}}$ and a magnetic surface current density
$c\mathbf{M}\times\hat{\mathbf{n}}$. 
Relativistic invariance is enforced by defining the antisymmetric dipolar
density tensor $D_{\alpha\beta}$, with its spatial components obtained from the
magnetization density by $D_{ij}=\epsilon_{ijk}M_k$
and its temporal components given by the electric polarization, $D_{0k}=-D_{k0}=P_k$. The charges and currents associated with $\mathbf{P}$ and $\mathbf{M}$ are
encoded in the dipolar four current $j_\mathrm{dip}^\mu = c\partial_\nu D^{\mu\nu}$,
which like the free charge four current $j^\mu$,
is conserved: $\partial_\mu\partial_\nu D^{\mu\nu}=0$.
We work in Gauss units, the metric tensor is
$\eta^{\mu\nu}=\mathrm{diag}(-1,1,1,1)$
and $c$ is the speed of light in vacuum.  Maxwell equations  are
\begin{equation}
\label{Maxwell}
\partial_\nu F^{\mu\nu} = \frac{4\pi}{c}(j^\mu+j_\mathrm{dip}^\mu)\ \ ,
\end{equation}
where $F^{\mu\nu}$ is the electromagnetic field tensor. Defining the tensor of
magnetizing field $\mathbf{H}$ and electric displacement $\mathbf{D}$ through
$H^{\mu\nu}=F^{\mu\nu}-4\pi D^{\mu\nu}$, the field equations become
$\partial_\nu H^{\mu\nu}=4\pi c^{-1} j^\mu$.

Let us first consider briefly the case with vanishing $\mathbf{P}$ and $\mathbf{M}$. In this case Maxwell's equations read $\partial_\nu F^{\mu\nu}=4\pi c^{-1} j^\mu$. The force
density on the free charges is a four vector given by
$f^\mu_\mathrm{ch} = \frac{1}{c} F^\mu_{\ \nu}j^\nu$. Consider now the gauge
invariant symmetric tensor
\begin{equation}
\label{Standard}
T_{\mathrm S}^{\mu\nu}=
 -\frac{1}{16\pi}\eta^{\mu\nu}F^{\alpha\beta}F_{\alpha\beta}
+\frac{1}{4\pi}F^\mu_{\ \alpha}F^{\nu\alpha} \ .
\end{equation}
The relation 
\begin{equation}
\label{FieldForce}
\partial_\nu T_{\mathrm S}^{\mu\nu} = -\frac{1}{c} F^\mu_{\ \nu}j^\nu=
 -f^\mu_\mathrm{ch} 
\end{equation}
is an identity which holds for every solution of Maxwell equations. One is allowed to identify $T_{\mathrm S}^{\mu\nu}$ as the
energy-momentum tensor of the electromagnetic
field and to interpret the right hand side of (\ref{FieldForce}) as the force the matter exerts on the field. In particular Newton's action-reaction law holds. 

Consider now the case with non-vanishing $D^{\mu\nu}$. Although some authors suppose that the force on matter is of the form $f^\mu_\mathrm{ch}$ with $j^\mu$ substituted by $j^\mu+j_\mathrm{dip}^\mu$ (See for example \cite{ObkYH2003,Obk2008}) it is easy to be convinced that this is  not the case. We obtain the force density expression assuming that 1)The total force on a piece of material is the sum of the forces on each element of the piece. 2)The force on an element equals the force on the dipoles $d\mathbf{m}=\mathbf{M}dV$ and $d\mathbf{d}=\mathbf{P}dV$.  The force on a magnetic dipole $\mathbf{m}$ is known to be \cite{JacJ1998} $\mathbf{F}_{\mathrm{dip}}=\nabla(\mathbf{B}\cdot\mathbf{m})$. The power transferred to matter is  $\frac{dW}{dt}= -\frac{\partial\mathbf{B}}{\partial t}\cdot\mathbf{m}$. Using that in this case $\mathbf{P}=0$, the relativistic force density on the microscopic dipoles 
is $f^\mu_{\mathrm{dip}}={2}^{-1}D_{\alpha\beta}\partial^\mu F^{\alpha\beta}$. For non vanishing $\mathbf{P}$
and $\mathbf{M}$, by relativistic invariance the total force density four-vector is
\begin{equation}
\label{force-density}
f^\mu =f^\mu_\mathrm{ch}+f^\mu_\mathrm{dip}=\frac{1}{c} F^{\mu\nu} j_\nu +
 \frac{1}{2}D_{\alpha\beta}\partial^\mu F^{\alpha\beta}\ .
\end{equation}
A related expression is discussed in \cite{deGrootS1972}. The energy-momentum tensor of matter  satisfies,
 \begin{equation}
 \label{Tmatter}
  \partial_\nu T_{\mathrm{matter}}^{\mu\nu}=f^\mu \ .
\end{equation}
Equation (\ref{FieldForce}) is an identity which follows from Maxwell's equations.  Using (\ref{Maxwell}) in this case  we can write directly the new identity
\begin{eqnarray}
\label{Identity}
\partial_\nu T_{\mathrm S}^{\mu\nu} =-\frac{1}{c} F^\mu_{\ \nu}(j^\nu+j_\mathrm{dip}^\mu)=-\frac{1}{c} F^\mu_{\ \nu}(j^\nu+c\partial_\alpha D^{\nu\alpha}).
\end{eqnarray}
The right hand side of
(\ref{Identity}) is not minus the total force on the matter
(\ref{force-density}) and the  identification of
$T_{\mathrm S}^{\mu\nu}$ as the energy-momentum of the field does not hold. One
regains a clear physical interpretation  by defining 
\begin{equation}
\label{energy-momentum-tensor}
T^{\mu\nu}_{\mathrm{FK}}=
-\frac{1}{16\pi}F_{\alpha\beta}F^{\alpha\beta}\eta^{\mu\nu}
+\frac{1}{4\pi}F^{\mu}_{\ \alpha}H^{\nu\alpha}\ .
\end{equation}
which after a simple manipulation using Bianchi's identity, Eq. (\ref{Identity})
is shown to satisfy
\begin{equation}
\label{energy-momentum-equation}
\partial_\nu T_{\mathrm{FK}}^{\mu\nu} = -f^\mu\ .
\end{equation}
Newton's third law between matter and field is recovered if one identify $T^{\mu\nu}_{\mathrm{FK}}$ as the kinetic energy-momentum tensor of the electromagneticfield. Of course different energy-momentum tensors may be used for particular
purposes, but $T^{\mu\nu}_{\mathrm{FK}}$ is the one that should be used to discuss exchange
of linear momentum between matter and the electromagnetic field because it is
Newton's third law which guarantees the conservation of the total energy-momentum tensor. With this tensor the energy density is
\begin{equation}
\label{energy-density}
u = T_{\mathrm FK}^{00} = \frac{1}{8\pi}(E^2+B^2)+\mathbf{E}\cdot\mathbf{P}\ \ ,
\end{equation}
and  Poynting vector and the momentum density are 
\begin{equation}\mathbf{S} = c T_{\mathrm FK}^{0i}=\frac{c}{4\pi}\mathbf{E}\times\mathbf{H} \ \ ,\ \ \mathbf{g} = c^{-1}  T_{\mathrm FK}^{i0}=
\frac{1}{4\pi c}\mathbf{D}\times\mathbf{B}. 
\end{equation} 
Maxwell's stress tensor is
\begin{equation}
\label{Maxwell-tensor}
 T_{\mathrm FK}^{ij}  = \frac{1}{8\pi}(E^2+B^2)\delta_{ij} 
-\mathbf{B}\cdot\mathbf{M}\delta_{ij} -\frac{1}{4\pi}(E_i D_j +H_i B_j) \ \ .
\end{equation}
The obtained tensor is different to Minkowski's and Abraham's tensors. Minkowski's tensor in our notation reduces to
\begin{eqnarray} 
\label{Mikowski-tensor}
T^{\mu\nu}_{\mathrm{Min}}= T^{\mu\nu}_{\mathrm{FK}}
+\frac{1}{4}F^{\alpha\beta}D_{\alpha\beta}\eta^{\mu\nu}\ .
\end{eqnarray} 
It differs from $T_{\mathrm{FK}}$ by diagonal terms.  Poynting's vector
and the momentum density are the same for both tensors but the classical Minkowski or Poynting energy density  $u_\mathrm{Min}=(\mathbf{E}\cdot\mathbf{D}+\mathbf{B}\cdot\mathbf{H})/8\pi$ \cite{Poynting}   is different from the expression (\ref{energy-density}). The diagonal terms of the Maxwell tensor are also different. 
Abraham's tensor cannot be written in covariant form. This fact was shown in Ref.\cite{VeselagoShchavlev2010} by an explicit computation and has also a simple demonstration because there is a unique four-tensor that has some
particular temporal row inevry reference frame and Abraham's and Minkowski's two indices objects have the same
temporal row. Our tensor is related but different to the  one obtained by de Groot and Suttorp in a particular case \cite{deGrootS1972}

The non-symmetric part of  $T^{\mu\nu}_{\mathrm{FK}}$ has  to be interpreted in view of equation (\ref{spin-eq}) as a dipolar torque density
\begin{equation}
\label{dip-torque}
\tau^{\mu\nu}_{\mathrm{dip}}=
D^{\mu\beta}F^\nu_{\ \beta} - D^{\nu\beta}F^\mu_{\ \beta}
\end{equation} 
Inspecting its components one observes that indeed the spatial part is given by
\begin{equation}
\label{torque-space}
(\mathbf{\tau}_{\mathrm{dip}})_k=\frac{1}{2}\epsilon_{ijk}\tau^{ij}_\mathrm{dip}
= (\mathbf{P}\times\mathbf{E} + \mathbf{M}\times\mathbf{B})_k \ ,
\end{equation} 
which is the expected  torque that the field should exert on magnetic and electric dipoles. The temporal part is,
\begin{equation}
\label{torque-time}
\tau^{0k}_{\mathrm{dip}}= (-\mathbf{P}\times\mathbf{B}
+\mathbf{M}\times\mathbf{E})^k \ .
\end{equation} 
and as we discuss in the following example plays an important role in disentangling the paradoxes of Balazs construction.

The best test for the energy-momentum tensor and the force density presented in
this letter is to compute the momentum and energy exchange between a packet of
electromagnetic waves and a dielectric medium.
Suppose that the region  $x>0$ is filled
by a non-dispersive material with dielectric constant $\epsilon$
and magnetic permeability $\mu$. A packet of linearly polarized
plane waves approaches the $yz$ surface traveling in the $x$ direction.
Its electric field is
\begin{equation}
\mathbf{E}_1(x,y,z,t) = E_1 g(t-x/c)\theta(-x)\hat{y} \ .
\end{equation}
$E_1$ is an amplitude, $\theta$ is the Heaviside step function and $g(t)$ is
a dimension-less well-behaved but otherwise arbitrary function that vanishes
for $t<0$ and $t>T$. At the surface of
the material $x=0$ the packet is reflected and transmitted. The reflected and
transmitted packets  are
\begin{eqnarray}
\mathbf{E}_2(x,y,z,t) &=& E_2 g(t+x/c)\theta(-x)\hat{y} \ ,\\
\mathbf{E}_3(x,y,z,t) &=& E_3 g(t-x/v)\theta(x)\hat{y} \ ,
\end{eqnarray}
where the speed of light in the material is $v=c/n$ with $n=\sqrt{\epsilon\mu}$.
For $t<0$ only the incident packet is present, for $t>T$  the
reflected one is in $x<0$ and the transmitted one is in $x>0$. For $0<t<T$ the
three packets are touching the surface $x=0$.
The corresponding magnetic fields of the three packets are
\begin{eqnarray}
\mathbf{B}_1 &=& B_1 g(t-x/c)\theta(-x)\hat{z}\ ,\\
\mathbf{B}_2 &=& B_2 g(t+x/c)\theta(-x)\hat{z}\ ,\\
\mathbf{B}_3 &=& B_3 g(t-x/v)\theta(x)\hat{z}\ .
\end{eqnarray}
Using Maxwell's equations the magnetic amplitudes are 
\begin{equation}
B_1 = E_1\ ,\qquad B_2 = -E_2\ , \qquad B_3 = \sqrt{\epsilon\mu}E_3\ .
\end{equation}
By the continuity conditions at $x=0$
\begin{eqnarray}
E_2 = \frac{1-\sqrt{\epsilon/\mu}}{1+\sqrt{\epsilon/\mu}}E_1\ ,\ \  
E_3 = \frac{2}{1+\sqrt{\epsilon/\mu}}E_1 \ .
\end{eqnarray}
For $t<0$ the energy  of a cylindrical piece of
the incident packet with axis parallel to $x$ and cross section  $A$ is,
\begin{eqnarray}
U_1 &=& \int T^{00}_{\mathrm{S}}(1)\,dV
=\frac{Ac\bar{T}}{4\pi}E_1^2
\end{eqnarray}
with
\begin{equation}
\bar{T} = \int_0^T g(t)^2\,dt \ .
\end{equation}
The momentum of the incident wave-packet is
\begin{equation}
\mathbf{p}_1=\int\mathbf{g}(1)\,dV=
\int c^{-1} T^{i0}_\mathrm{S}(1)\hat{\mathbf{e}}_i\,dV  =
 \frac{U_1}{c}\hat{x} \ .
\end{equation}
For the reflected packet  ($t>T$) the energy and momentum are
\begin{equation}
U_2 = \frac{Ac\bar{T}}{4\pi}E^2_2\ ,\quad  
\mathbf{p}_2 = \int \mathbf{g}(2)\,dV = -\frac{U_2}{c}\hat{x}\ .
\end{equation}

The energy and momentum  transferred to the $x>0$ side of the
space are
\begin{eqnarray}
U_1-U_2 &=& \frac{Ac\bar{T}}{4\pi}(E^2_1-E^2_2)=
\frac{Ac\bar{T}}{4\pi}E^2_3\sqrt{\epsilon/\mu}\\
\mathbf{p}_1-\mathbf{p}_2&=&\frac{A\bar{T}}{4\pi}(E^2_1+E^2_2)\hat{x}
=\frac{A\bar{T}}{8\pi}E^2_3(1+\epsilon/\mu)\hat{x}\ .\ \;
\end{eqnarray}

The EM energy and momentum of the transmitted packet are
\begin{eqnarray}
U_3 &=& \int T^{00}_{\mathrm{FK}}(3)\,dV =\frac{Ac\bar{T}}{8\pi\sqrt{\epsilon\mu}}
E^2_3(\epsilon\mu+2\epsilon-1) \ ,\\
\mathbf{p}_3 &=& \int \mathbf{g}(3)\,dV=
 \frac{A\bar{T}v}{4\pi c}E^2_3\epsilon\sqrt{\epsilon
\mu}\hat{x} \ .
\end{eqnarray}
Using (\ref{force-density}) the power on the matter at time $t$ is obtained
\begin{eqnarray}
\dot{W}&=&c\int f^0 dv=-\int(\mathbf{P}\cdot\dot{\mathbf{E}}+\mathbf{M}\cdot\dot{\mathbf{B}})dV\nonumber\\
&=& - \frac{Ac}{8\pi\sqrt{\epsilon\mu}}E^2_3(\epsilon\mu -1)g(t)^2\ .
\end{eqnarray}
Integrating the time the work done on matter is
\begin{equation}
W = -\frac{Ac\bar{T}}{8\pi\sqrt{\epsilon\mu}}E^2_3(\epsilon\mu-1) \ .
\end{equation}
This work changes the energy of the matter where the wave-packet is located, so
it has to be added to the EM energy in order to obtain the total transmitted energy  $U^\prime_3 = U_3+W$. Energy conservation is satisfied
$U^\prime_3 =U_1-U_2$. It is easy to see that $U^\prime_3$ is the energy 
of the transmitted packet computed with $u_\mathrm{Min}$. Note also that
 $\mathbf{p}_3=c^{-1}U^\prime_3 n\hat{x}$
as would be expected for Minkowski's momentum.

To verify momentum conservation one has to compute
the impulse on matter. The force on matter has a volume component
given by (\ref{force-density}) and a surface component due
to the discontinuity at $x=0$. The volume component is 
\begin{eqnarray}
\label{force-V}
\mathbf{F}_{\mathrm V} &=&\int (P_i\nabla E_i+M_i\nabla B_i)dV\nonumber\\
&=& \frac{A}{8\pi}\int_0^\infty[(\epsilon-1)\partial_x E^2+(1-1/\mu)
\partial_x B^2]dx \,\hat{x}\nonumber\\
&=&-\frac{AE^2_3}{8\pi}(\epsilon\mu-1)g(t)^2\hat{x} \ .
\end{eqnarray}
The surface component of the force at $x=0$ is equal to the momentum flux
exiting the vacuum side minus the momentum flux entering the matter side.
That is
\begin{equation}
\mathbf{F}_{\mathrm S}=
A(T^{11}_{\mathrm{S}}(-)-T^{11}_{\mathrm{FK}}(+))\hat{x} \ .
\end{equation}
Using (\ref{Standard}) and (\ref{energy-momentum-tensor} )
\begin{eqnarray}
T^{11}_{\mathrm{S}}(-)-T^{11}_{\mathrm{FK}}(+)
=\frac{\epsilon E^2_3}{8\pi}(1/\mu+\mu-2)g(t)^2 \ .
\end{eqnarray}
Therefore the total force is
\begin{equation}
\label{force-VS}
\mathbf{F}=\mathbf{F}_{\mathrm V}+\mathbf{F}_{\mathrm S}=
\frac{AE^2_3}{8\pi}(1+\epsilon/\mu- 2\epsilon)g(t)^2\hat{x} \ .
\end{equation}
We note that if diamagnetism does not prevail the wave packet pulls the dielectric. The impulse is
\begin{equation}
\label{Impulse}
\mathbf{I} = \int \mathbf{F}\,dt=
\frac{A\bar{T}E^2_3}{8\pi}(1+\epsilon/\mu-2\epsilon)
\hat{x} \ .
\end{equation}
The total momentum transferred to  $x>0$ for $t>T$ is
\begin{equation}
\mathbf{I}+\mathbf{p}_3=\frac{A\bar{T}E^2_3}{8\pi}(1+\epsilon/\mu)\hat{x}
=\mathbf{p}_1-\mathbf{p}_2
\end{equation}
as it should be.

Let us turn to the motion of the center of mass of the system. It is convenient to separate the electromagnetic and matter contributions to the center of mass and write
\begin{equation}
\mathbf{X}_\mathrm{T}(t)=\mathbf{X}_\mathrm{FT}(t)+\mathbf{X}_\mathrm{MT}(t)
\end{equation}
When the wave is moving towards the dielectric there is no spin contribution to the center of mass and spin an we may write 
\begin{equation}
 \dot{\mathbf{X}}_\Theta= \dot{\mathbf{X}}_\mathrm{T}= \dot{\mathbf{X}}_\mathrm{FT}=\frac{c^2p_1}{U_1}\hat{x}\  \  ,\ \ t<0\ .
\end{equation}
The position of the center of mass of the transmitted wave-packet  for $t>T$ is
\begin{equation}
\label{CenterOfMass}
\mathbf{X}_\mathrm{FT}(t)=
\frac{1}{U_3}\int xu\,dV\,\hat{x}
=\frac{1}{v\bar{T}}\int x g(t- x/v)^2 \,dx\,\hat{x}\ .
\end{equation}
It immediately follows that $\mathbf{X}_\mathrm{FT}(t) = \mathbf{X}_\mathrm{FT}(0) + tv\hat{x}$. The center of mass  velocity of this packet  $\dot{\mathbf{X}}_\mathrm{FT}=v\hat{x}$  is in this case indeed constant and  can be easily expressed as
\begin{equation}
\label{CMvelocity}
\dot{\mathbf{X}}_\mathrm{FT}=\frac{1}{U^\prime_3}\int \mathbf{S}\,dV=
\frac{1}{U^\prime_3}\int T^{oi}_\mathrm{S}(1)\hat{\mathbf{e}}_i\,dV \ ,
\end{equation}
but the strong CMMT does not hold
($\mathbf{p}_3\not=c^{-2}U^\prime_3\dot{\mathbf{X}}$) since
$\mathbf{g}\not=c^{-2}\mathbf{S}$. If the momentum of the transmitted
wave-packet were Abraham's the CMMT would be satisfied but the momentum
conservation would be lost. Let us then compute the spin contribution. After the wave has penetrated the dielectric, the center of mass of matter satisfies Newton's second law  $m\dot{\mathbf{X}}_\mathrm{MT}=\mathbf{I}$ where $m$ is the mass of the dielectric block and $\mathbf{I}$ is the impulse computed in (\ref{Impulse}). The spin density has contributions from matter and field and satisfies equation (\ref{spin-eq}).  The separation of these contributions is an difficult and interesting problem which is not necessary to discuss here. Since the matter contribution to the energy-momentum tensor is symmetric, using (\ref{dip-torque}) we have 
 \begin{equation}
\label{torque-spin}
\partial_\alpha S^{\mu\nu\alpha} = \tau^{\mu\nu}_{\mathrm{dip}} \ .
\end{equation}
with $\tau^{\mu\nu}_{\mathrm{dip}}$ given by (\ref{torque-space}) and (\ref{torque-time}). Focusing in the temporal components which are the ones that contribute  to (\ref{spincenter})  we have,
\begin{eqnarray}
\label{spintorque}
\partial_\alpha S^{0k\alpha}&=&(-\mathbf{P}\times\mathbf{B}
+\mathbf{M}\times\mathbf{E})^k \nonumber=-\frac{\mu\epsilon-1}{4\pi\mu}(\mathbf{E}\times\mathbf{B})^k
\end{eqnarray}
where we use the constitutive equations ${4\pi}\mathbf{P}= (\epsilon-1)\mathbf{E}\  ,\  {4\pi\mu}\mathbf{M}=(\mu-1)\mathbf{B}\ $.
Now, spin transport in this system is due by the drift, $S^{0ki}=S^{0k0}v^i_m$ with $v^i_m$ the matter velocity which in this case vanishes. Then $\partial_iS^{0ki}=0$ and using that the right hand side of (\ref{spintorque}) points in the $x$ direction we have
\begin{equation}
 \partial_0S^{010}=-\frac{(\mu\epsilon-1)\sqrt{\epsilon\mu}}{4\pi\mu}g^2(t-x/v)E^2
\end{equation}
Integrating in space the spin term which appear in equation (\ref{CMSMT})  is for $t>T$
\begin{equation}
\frac{\partial}{\partial t}S^{010}=-\frac{AcE^2\bar{T}(\mu\epsilon-1)}{4\pi\mu}=-\frac{U_1\dot{X}^1_\mathrm{S}}{c}\ .
\end{equation}
Taking all together, for $t>T$ we verify  that for $t>T$, 
\begin{eqnarray}
 \dot{X}^1_\Theta&=&-\frac{U_2c+(U_3+W)v+I}{U_1c^2}+\frac{\dot{X}^1_\mathrm{S}}{U_1c^2}\nonumber\\&=&\frac{A\bar{T}E_1^2}{4\pi}=\frac{c^2p_1}{U_1}
\end{eqnarray}
as requested by the improved  theorem (\ref{CMSMT}).

\section{Conclusion}

Using relativistic invariance and Maxwell equations  we deduce an invariant
expression of the force density that the electromagnetic field exerts on
dipolar matter (\ref{force-density}). Imposing Newton's third law between the
field and matter, we construct the kinetic energy-momentum tensor of the
electromagnetic field in matter $T^{\mu\nu}_{\mathrm{FK}}$. Our result differs
from both Minkowski and Abraham proposals but settles the Minkowski-Abraham
controversy about the momentum density in favor of the former. The energy
density obtained is not Poynting's classical expression but energy conservation
is assured by the power contribution of the dipolar term in
Eq.(\ref{force-density}). 

We use force density and $T^{\mu\nu}_{\mathrm{FK}}$ to verify energy and momentum conservation in the interaction of a  packet of electromagnetic waves with a dielectric medium. We show that in this system the CMMT does not hold but the modified equation (\ref{CMSMT}) is satisfied with a non trivial contribution of the temporal spin. 

We have shown, in opposition to the argument of Balazs \cite{Balazs1953}, that for $n>1$  the wave packet pulls the material when it enters a medium (See Eq.(\ref{force-VS})).
Experimental support to this result was reported in \cite{Campbell}. Since
there has been some perplexity about this possibility, we note that it has a
very simple physical explanation. Dielectric and paramagnetic materials are
attracted while diamagnetic materials are repelled in the direction to high
field regions, so when the wave packet is entering the medium it pulls the
material unless diamagnetism prevails.  For the same reason when the wave
leaves, it drags the block. 

In general Minkowski's tensor is not particularly useful but for a material 
with non-dispersive linear polarizabilities ($D_{\alpha\beta}=\chi_{\alpha\beta\mu\nu}F^{\mu\nu}$), such as the one discussed above, it may be interpreted as the energy-momentum tensor
of the electromagnetic field plus the fraction of the energy of the matter that
is due to the polarizations. Nevertheless its divergence is not the reaction of the force acting on the matter. 

We also want to mention that the expression for $T^{\mu\nu}_{\mathrm{FK}}$ may also be obtained starting from the microscopic equations and using and averaging procedure\cite{MedandSc} or using Noether's theorem within the Lagrangian formalism  \cite{MedandSd}.

\end{document}